\documentclass[11pt, a4paper]{article}
\pdfoutput=1

\usepackage{amsmath}
\usepackage{amsfonts}
\usepackage{amssymb}
\usepackage{graphicx}
\usepackage{mathrsfs}
\usepackage{latexsym}
\usepackage{color}
\usepackage{cite}
\usepackage{slashed, cancel}
\usepackage{hyperref}
\usepackage{datetime}

\numberwithin{equation}{section}

\hypersetup{colorlinks, citecolor=bluscuro, linkcolor=bluscuro, urlcolor=bluscuro}
\definecolor{rossos}{rgb}{0.8,0.2,0.3}
\definecolor{bluscuro}{rgb}{0.15, 0.2, .85}
\definecolor{bluchiaro}{cmyk}{1,.3,0.,0.1}

\makeatother   
\newcommand{\GeV}{{\rm \,GeV}}
\newcommand{\TeV}{{\rm TeV}}

\def\de{\textrm{d}}

 \def\be   {\begin{equation}}   \def\ee   {\end{equation}}
 \def\ba   {\begin{array}}      \def\ea   {\end{array}}
 \def\bea  {\begin{eqnarray}}   \def\eea  {\end{eqnarray}}
 \def\bean {\begin{eqnarray*}}  \def\eean {\end{eqnarray*}}

\begin{document}

%
\begin{flushright} 

\end{flushright}

\vspace{1cm}
\begin{center}

{\LARGE \textbf {
On the Validity of the Effective Field Theory
\\
[0.01cm]
 for Dark Matter Searches at the LHC
}}
\\ [1.5cm]

{\large
\textsc{Enrico Morgante}\footnote{\texttt{enrico.morgante@unige.ch}}
}
\\[1.5cm]

\large{
\textit{D\'epartement de Physique Th\'eorique, Universit\'e de Gen\`eve\\
and Centre for Astroparticle Physics (CAP),\\
24 quai E. Ansermet, CH-1211 Geneva, Switzerland}
}
\end{center}

\vspace{1cm}

\begin{center}
\textbf{Abstract}
\begin{quote}
We review the limitations to the use of the effective field theory approach to study dark matter at the LHC.
Due to the high energy reach, the low energy description breaks down, and may lead to incorrect results.
The use of simplified models is suggested.

\vspace{1cm}

Prepared for the Proceedings of \emph{Incontri di Fisica delle Alte Energie 2014}, Laboratori Nazionali del Gran Sasso \& Gran Sasso Science Institute, 09-11/04/2014.

\end{quote}

\end{center}

\def\thefootnote{\arabic{footnote}}
\setcounter{footnote}{0}
\pagestyle{empty}

\newpage
\pagestyle{plain}
\setcounter{page}{1}

\section{Introduction}
%
%

To detect and measure the properties of the Dark Matter (DM) particle is one of the primary goals of the Large Hadron Collider (LHC).
Up to now, there has been no evidence of its discovery, and the upcoming $14\TeV$ data will be crucial to improve our knowledge.

One of the most promising channels for DM searches at the LHC is the so-called \emph{mono-jet} channel,
in which a couple of invisible DM particles is produced in association with a hadronic jet.
The signature in the collider setup would then be a single jet plus missing energy in the transverse plane.
To obtain from data interesting constraints on its properties, it's very important to find a model-independent way to parametrize the interaction between the DM particle and the colliding partons.
The simplest possibility is to write down a list of effective operators that are generated by integrating out heavy mediators.
This approach has three advantages:
first, the same operator can be constrained also by direct and indirect searches;
second, only the DM particle is added to the SM, making computations particularly easy;
third, all the information is embedded in two parameters (the mass of the DM and the energy scale of the operator) simplifying the parameter space to constrain.
For these reason, effective operators have been used by all phase 1 LHC analysis, giving constraints competitive with direct detection ones.

In this talk, based on the analysis we performed in Refs.\cite{Busoni:2013lha,Busoni:2014sya}, we want to highlight the limitations of the EFT approach due to the high energy reach of the LHC,
and suggest a possible way out that should be followed in the analysis of the next LHC run.

\section{Domain of validity of EFT}

Effective operators that couple SM particles to the DM can be obtained, in a generic high energy new physics model, as a low energy approximation:
when the typical momentum exchanged in the process is much lower than the mass of the mediating particles, the effective description applies.
%
%

Suppose, for example, that the process of DM production with the emission of a gluon or a photon happens through the exchange of a virtual mediator, with the gluon/photon radiated from the initial state partons.
Then, the effective description consists in expand at zeroth order the denominator in the propagator for small transferred momentum:
\be
\label{eq:expansion}
\frac{g_q g_\chi}{Q_{\rm tr}^2-M^2}
\simeq -\frac{g_q g_\chi}{M^2}
\equiv -\frac{1}{\Lambda^2}\,,
\ee
where $g_q,g_\chi$ are the coupling constants of the mediator with the quarks and the DM, respectively.

From Eq.\ref{eq:expansion} is clear that the effective field theory approach is only valid if $Q_{\rm tr}^2 \ll M^2$,
\emph{i.e.} $Q_{\rm tr}^2 \ll \Lambda^2$ in the EFT language, assuming couplings of order 1.
Given this condition, we can define a quantity (which we call $R_\Lambda^{\rm tot}$) that quantifies the validity of the EFT approach for a given experimental setup.
$R_\Lambda^{\rm tot}$ is defined as following:
\be
R_\Lambda^{\rm tot}\equiv\frac{\sigma_{\rm eff}\vert_{Q_{\rm tr}<\Lambda}}{\sigma_{\rm eff}}
=\frac{\int_{p_{\rm T}^{\rm min}}^{1 \TeV}\de p_{\rm T}\int_{-2}^2\de \eta
\left.\dfrac{\de^2\sigma_{\rm eff}}{\de p_{\rm T}\de\eta}\right\vert_{Q_{\rm tr}<\Lambda}}
{\int_{p_{\rm T}^{\rm min}}^{1 \TeV}\de p_{\rm T}\int_{-2}^2\de \eta
\dfrac{\de^2\sigma_{\rm eff}}{\de p_{\rm T}\de\eta}},
\label{ratiolambdatot}
\ee
where $\sigma_{\rm eff}$ is the cross section obtained in the effective theory,
and the notation $\vert_{Q_{\rm tr}<\Lambda}$ indicates that, in the numerator, the integrals over the jet variables $p_{\rm T},\eta$ and over the PDFs are performed only on the region of the phase space in which the condition $Q_{\rm tr}<\Lambda$ holds.
The limits in the integrals are chosen to match the standard choices made by the experimental collaboration.

With respect to the other quantities introduced in \cite{Busoni:2013lha}, $R_\Lambda^{\rm tot}$ has two main advantages:
first, it has a direct interpretation as the fraction of events in an experimental setup in which the EFT description is valid.
Second, it is independent on the specific UV completion of the theory.

\begin{figure}[htb]
\centering
\includegraphics[scale=.5]{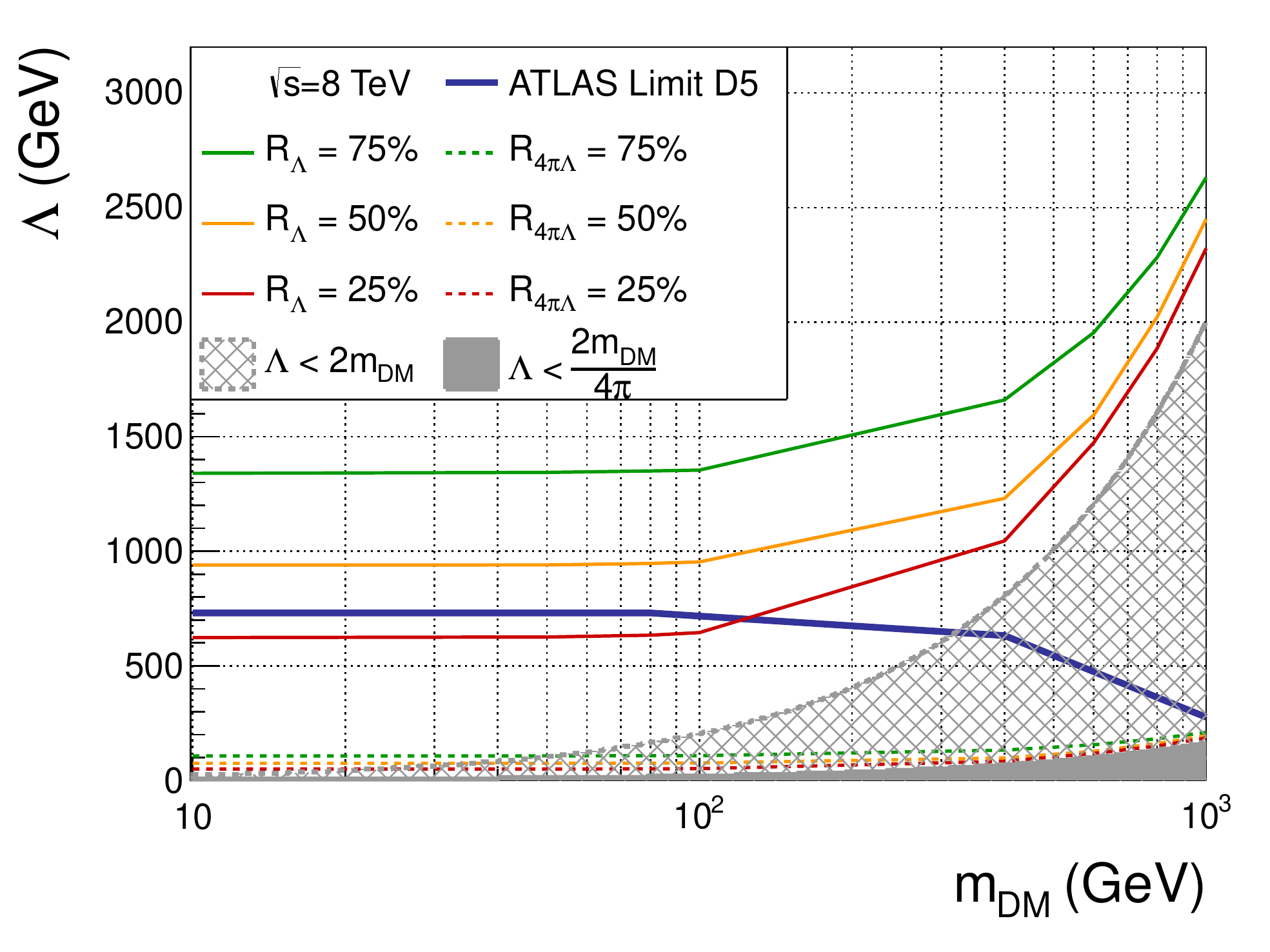}
\caption{ 
\emph{{\small
Contours for the ratios $R_{\Lambda}^{\rm tot}$ and $R_{4\pi\Lambda}^{\rm tot}$ on the plane
$(m_{\rm DM}, \Lambda)$. We set $\sqrt{s}=8 \TeV,
|\eta|\leq 2$ and $p_{\rm T}^{\rm min}=120 \GeV$, taken from \cite{Busoni:2014sya}. 
We have also shown the contour corresponding to $\Lambda<2 m_{\rm DM}$ and $\Lambda<m_{\rm DM}/(2\pi)$
which are often used as a benchmark for the validity of the EFT.
Blue line: ATLAS limit \cite{ATLAS}.
}}}
\label{fig:RLambdatot}
\end{figure}

In Fig.\ref{fig:RLambdatot} different contours for $R_\Lambda^{\rm tot}$ in the plane $m_{\rm DM} \textsl{ vs. }\Lambda$ are shown
for the operator $(\bar q \gamma^\mu q)(\bar\chi\gamma_\mu\chi)/\Lambda^2$, usually referred to as D5. We have assumed for simplicity that the DM is a Dirac fermion.
The first thing that we learn from Fig.\ref{fig:RLambdatot} is that the limit on the operator D5 put by ATLAS collaboration in \cite{ATLAS} falls in the region where $R_\Lambda^{\rm tot}$ is far from the value of $1$ needed for the validity of EFT.
This means that, assuming the couplings are of order $1$, in the majority of the events considered in the analysis the EFT approach was non applicable, and the actual limit obtained considering only a fraction $R_\Lambda^{\rm tot}$ of the events would be considerably weaker \cite{Busoni:2014sya}.

The second important point is that, if we assume the couplings to be of order $\mathcal{O}(4\pi)$, the validity of the effective approach is restored:
this can be seen by looking at the contours for the quantity $R_{4\pi\Lambda}^{\rm tot}$, defined in the same way of $R_\Lambda^{\rm tot}$ but with the condition $Q_{\rm tr}<4\pi\Lambda$ in the numerator (dashed lines in Fig.\ref{fig:RLambdatot}).
We see that the effective approach is only valid if the coupling constants are greater than some value that depends on the specific operator and on the details of the experimental setup.

%

\section{Conclusions and possible directions}


The limitation that we highlight makes very urgent the need for a more robust description of the interactions between DM and SM particles that, on the other hand, does not have the complicated structure of a full new physics theory like SUSY.
One possible solution is the use of the so called \emph{simplified models} \cite{Abdallah:2014dma}:
these are toy models in which we add to the SM only two more particles, the DM and one new mediator, and a few new parameters, \emph{i.e.} the masses and couplings of these particles.
We then use tree level approximation to compute all the relevant transition amplitudes.

This approach has three advantages:
first, it doesn't suffer from the limitations of the EFT;
second, it grasps relevant physical features like resonances in the DM production cross section;
third, the interesting complementarity with direct searches of the mediator can be exploited to further restrict the parameter space (see \emph{e.g.}\cite{Papucci:2014iwa}).
We therefore suggest that, in the future, simplified models will be used to analyze experimental data instead of EFT.

\section*{Acknowledgments}
I greatfully thank Giorgio Busoni, Andrea De Simone, Johanna Gramling, Thomas Jacques and Toni Riotto.
I also thank all the organizing commitee of IFAE 2014 for the great opportunity of being immersed in such an exciting environment in the beautiful landscape of Abruzzo mountains.

\end{document}